\pdfoutput=1

\documentclass[12pt,a4paper]{article}

\usepackage{ifthen} 
\newboolean{pdflatex}
\setboolean{pdflatex}{true} 

\newboolean{articletitles}
\setboolean{articletitles}{true} 

\newboolean{uprightparticles}
\setboolean{uprightparticles}{false} 

\newboolean{inbibliography}
\setboolean{inbibliography}{false} 

\usepackage{microtype}
\usepackage{lineno}  
\usepackage{xspace} 
\usepackage{caption} 

\usepackage{graphicx}  
\usepackage{color}
\usepackage{colortbl}
\graphicspath{{./figs/}} 

\usepackage{amsmath} 
\usepackage{amssymb}
\usepackage{amsfonts}
\usepackage{upgreek} 

\newcommand*\patchAmsMathEnvironmentForLineno[1]{%
\expandafter\let\csname old#1\expandafter\endcsname\csname #1\endcsname
\expandafter\let\csname oldend#1\expandafter\endcsname\csname
end#1\endcsname
 \renewenvironment{#1}%
   {\linenomath\csname old#1\endcsname}%
   {\csname oldend#1\endcsname\endlinenomath}%
}
\newcommand*\patchBothAmsMathEnvironmentsForLineno[1]{%
  \patchAmsMathEnvironmentForLineno{#1}%
  \patchAmsMathEnvironmentForLineno{#1*}%
}
\AtBeginDocument{%
\patchBothAmsMathEnvironmentsForLineno{equation}%
\patchBothAmsMathEnvironmentsForLineno{align}%
\patchBothAmsMathEnvironmentsForLineno{flalign}%
\patchBothAmsMathEnvironmentsForLineno{alignat}%
\patchBothAmsMathEnvironmentsForLineno{gather}%
\patchBothAmsMathEnvironmentsForLineno{multline}%
\patchBothAmsMathEnvironmentsForLineno{eqnarray}%
}

\usepackage{hyperref}    
\usepackage[all]{hypcap} 

\def\lhcb {\mbox{LHCb}\xspace}

\ifthenelse{\boolean{uprightparticles}}%
{

 \def\Pmu         {\ensuremath{\upmu}\xspace}

 \def\Ppi         {\ensuremath{\uppi}\xspace}

 \def\Pphi        {\ensuremath{\upphi}\xspace}

 \def\Ppsi        {\ensuremath{\uppsi}\xspace}

 \def\PDelta      {\ensuremath{\Delta}\xspace}                 
 \def\PXi      {\ensuremath{\Xi}\xspace}                 
 \def\PLambda      {\ensuremath{\Lambda}\xspace}                 
 \def\PSigma      {\ensuremath{\Sigma}\xspace}                 
 \def\POmega      {\ensuremath{\Omega}\xspace}                 
 \def\PUpsilon      {\ensuremath{\Upsilon}\xspace}                 
 
 \def\PB      {\ensuremath{\mathrm{B}}\xspace}                 
                  
 \def\PD      {\ensuremath{\mathrm{D}}\xspace}

 \def\PJ      {\ensuremath{\mathrm{J}}\xspace}                 
 \def\PK      {\ensuremath{\mathrm{K}}\xspace}

 \def\Pb      {\ensuremath{\mathrm{b}}\xspace}                 
 \def\Pc      {\ensuremath{\mathrm{c}}\xspace}                 
 \def\Pd      {\ensuremath{\mathrm{d}}\xspace}

 \def\Pi      {\ensuremath{\mathrm{i}}\xspace}

 \def\Pp      {\ensuremath{\mathrm{p}}\xspace}                 
 \def\Pq      {\ensuremath{\mathrm{q}}\xspace}                 
                  
 \def\Ps      {\ensuremath{\mathrm{s}}\xspace}

}
{

 \def\Pmu         {\ensuremath{\mu}\xspace}

 \def\Ppi         {\ensuremath{\pi}\xspace}

 \def\Pphi        {\ensuremath{\phi}\xspace}

 \def\Ppsi        {\ensuremath{\psi}\xspace}                 
                  
 \mathchardef\PDelta="7101
 \mathchardef\PXi="7104
 \mathchardef\PLambda="7103
 \mathchardef\PSigma="7106
 \mathchardef\POmega="710A
 \mathchardef\PUpsilon="7107
                  
 \def\PB      {\ensuremath{B}\xspace}                 
                  
 \def\PD      {\ensuremath{D}\xspace}

 \def\PJ      {\ensuremath{J}\xspace}                 
 \def\PK      {\ensuremath{K}\xspace}

 \def\Pb      {\ensuremath{b}\xspace}                 
 \def\Pc      {\ensuremath{c}\xspace}                 
 \def\Pd      {\ensuremath{d}\xspace}

 \def\Pi      {\ensuremath{i}\xspace}

 \def\Pp      {\ensuremath{p}\xspace}                 
 \def\Pq      {\ensuremath{q}\xspace}                 
                  
 \def\Ps      {\ensuremath{s}\xspace}

}

\def\mumu       {{\ensuremath{\Pmu^+\Pmu^-}}\xspace}

\def\quark     {{\ensuremath{\Pq}}\xspace}

\def\dquark    {{\ensuremath{\Pd}}\xspace}

\def\squark    {{\ensuremath{\Ps}}\xspace}

\def\cquark    {{\ensuremath{\Pc}}\xspace}
\def\cquarkbar {{\ensuremath{\overline \cquark}}\xspace}

\def\bquark    {{\ensuremath{\Pb}}\xspace}

\def\hadron {{\ensuremath{\Ph}}\xspace}
\def\pion   {{\ensuremath{\Ppi}}\xspace}
\def\piz    {{\ensuremath{\pion^0}}\xspace}

\def\pip    {{\ensuremath{\pion^+}}\xspace}
\def\pim    {{\ensuremath{\pion^-}}\xspace}
\def\pipm   {{\ensuremath{\pion^\pm}}\xspace}

\def\kaon    {{\ensuremath{\PK}}\xspace}
  \def\Kbar    {{\kern 0.2em\overline{\kern -0.2em \PK}{}}\xspace}

\def\Kp      {{\ensuremath{\kaon^+}}\xspace}
\def\Km      {{\ensuremath{\kaon^-}}\xspace}
\def\Kpm     {{\ensuremath{\kaon^\pm}}\xspace}

\def\KS      {{\ensuremath{\kaon^0_{\rm\scriptscriptstyle S}}}\xspace}

  \def\Dbar    {{\kern 0.2em\overline{\kern -0.2em \PD}{}}\xspace}
\def\D       {{\ensuremath{\PD}}\xspace}

\def\Dz      {{\ensuremath{\D^0}}\xspace}

\def\Dm      {{\ensuremath{\D^-}}\xspace}

\def\Dstar   {{\ensuremath{\D^*}}\xspace}

\def\Ds      {{\ensuremath{\D^+_\squark}}\xspace}

\def\B       {{\ensuremath{\PB}}\xspace}
\def\Bbar    {{\ensuremath{\kern 0.18em\overline{\kern -0.18em \PB}{}}}\xspace}

\def\Bz      {{\ensuremath{\B^0}}\xspace}

\def\Bd      {{\ensuremath{\B^0}}\xspace}
\def\Bs      {{\ensuremath{\B^0_\squark}}\xspace}

\def\Bdb     {{\ensuremath{\Bbar^0}}\xspace}

\def\jpsi     {{\ensuremath{{\PJ\mskip -3mu/\mskip -2mu\Ppsi\mskip 2mu}}}\xspace}

  \def\Y#1S{\ensuremath{\PUpsilon{(#1S)}}\xspace}

\def\proton      {{\ensuremath{\Pp}}\xspace}
\def\antiproton  {{\ensuremath{\overline \proton}}\xspace}

\def\Lz          {{\ensuremath{\PLambda}}\xspace}
\def\Lbar        {{\ensuremath{\kern 0.1em\overline{\kern -0.1em\PLambda}}}\xspace}

\def\Lb      {{\ensuremath{\Lz^0_\bquark}}\xspace}
\def\Lbbar   {{\ensuremath{\Lbar^0_\bquark}}\xspace}
\def\Lc      {{\ensuremath{\Lz^+_\cquark}}\xspace}

\newcommand{\decay}[2]{\ensuremath{#1\!\to #2}\xspace}         

\def\to                 {\ensuremath{\rightarrow}\xspace}

\def\CP                {{\ensuremath{C\!P}}\xspace}

\def\Vcd  {{\ensuremath{V_{\cquark\dquark}}}\xspace}

\def\Vcs  {{\ensuremath{V_{\cquark\squark}}}\xspace}

\def\AT#1     {\ensuremath{A_{\mathrm{T}}^{#1}}\xspace}           

\def\C#1      {\ensuremath{\mathcal{C}_{#1}}\xspace}                       
\def\Cp#1     {\ensuremath{\mathcal{C}_{#1}^{'}}\xspace}                    
\def\Ceff#1   {\ensuremath{\mathcal{C}_{#1}^{\mathrm{(eff)}}}\xspace}        
\def\Cpeff#1  {\ensuremath{\mathcal{C}_{#1}^{'\mathrm{(eff)}}}\xspace}       
\def\Ope#1    {\ensuremath{\mathcal{O}_{#1}}\xspace}                       
\def\Opep#1   {\ensuremath{\mathcal{O}_{#1}^{'}}\xspace}                    

\newcommand{\tev}{\ifthenelse{\boolean{inbibliography}}{\ensuremath{~T\kern -0.05em eV}\xspace}{\ensuremath{\mathrm{\,Te\kern -0.1em V}}}\xspace}
\newcommand{\gev}{\ensuremath{\mathrm{\,Ge\kern -0.1em V}}\xspace}
\newcommand{\mev}{\ensuremath{\mathrm{\,Me\kern -0.1em V}}\xspace}
\newcommand{\kev}{\ensuremath{\mathrm{\,ke\kern -0.1em V}}\xspace}
\newcommand{\ev}{\ensuremath{\mathrm{\,e\kern -0.1em V}}\xspace}
\newcommand{\gevc}{\ensuremath{{\mathrm{\,Ge\kern -0.1em V\!/}c}}\xspace}
\newcommand{\mevc}{\ensuremath{{\mathrm{\,Me\kern -0.1em V\!/}c}}\xspace}
\newcommand{\gevcc}{\ensuremath{{\mathrm{\,Ge\kern -0.1em V\!/}c^2}}\xspace}
\newcommand{\gevgevcccc}{\ensuremath{{\mathrm{\,Ge\kern -0.1em V^2\!/}c^4}}\xspace}
\newcommand{\mevcc}{\ensuremath{{\mathrm{\,Me\kern -0.1em V\!/}c^2}}\xspace}

\def\mum  {\ensuremath{{\,\upmu\rm m}}\xspace}

\def\invfb   {\ensuremath{\mbox{\,fb}^{-1}}\xspace}

\def\ps   {\ensuremath{{\rm \,ps}}\xspace}

\newcommand{\stat}{\ensuremath{\mathrm{\,(stat)}}\xspace}
\newcommand{\syst}{\ensuremath{\mathrm{\,(syst)}}\xspace}

\def\gsim{{~\raise.15em\hbox{$>$}\kern-.85em
          \lower.35em\hbox{$\sim$}~}\xspace}
\def\lsim{{~\raise.15em\hbox{$<$}\kern-.85em
          \lower.35em\hbox{$\sim$}~}\xspace}

\def\sPlot{\mbox{\em sPlot}}

\def\pt         {\mbox{$p_{\rm T}$}\xspace}

\def\evtgen     {\mbox{\textsc{EvtGen}}\xspace}

\def\geant      {\mbox{\textsc{Geant4}}\xspace}

\def\photos     {\mbox{\textsc{Photos}}\xspace}

\def\pythia     {\mbox{\textsc{Pythia}}\xspace}

\def\tell1  {TELL1\xspace}
\def\ukl1   {UKL1\xspace}

\usepackage{cite} 
\usepackage{mciteplus}

\usepackage{longtable} 
\usepackage{columntypes} 
\usepackage{enumerate}

\def\BFSystMC{\ensuremath{0.913\pm0.040}\xspace}
\def\BFSystTrigger{\ensuremath{1.000\pm0.010}\xspace} %

\def\BFSystLifetime{\ensuremath{1.000\pm0.001}\xspace}
\def\BFSystFit{\ensuremath{1.000\pm0.021}\xspace} 
\def\BFSystPID{\ensuremath{0.960\pm0.010}\xspace} 
\def\BFSystTot{\ensuremath{0.876\pm0.045}\xspace}
\newcommand{\IF}[4]{\ifthenelse{\equal{#1}{#2}}{#3}{#4}}%
\newcommand{\STAT}[1][E]{\IF{#1}{E}{\text{\:(stat)}}{}}
\newcommand{\SYST}[1][E]{\IF{#1}{E}{\text{\:(syst)}}{}}
\newcommand{\BFCorrected}[1][E]{\ensuremath{0.0824\pm0.0025\STAT[#1]\pm0.0042\SYST[#1]}\xspace} %
\def\RawRatio{\ensuremath{0.0940 \pm 0.0029}\xspace}

\def\RawLbpiCP{\ensuremath{(+7.9 \pm 2.2)\%}\xspace}
\def\RawLbKCP{\ensuremath{(+1.1 \pm 0.9)\%}\xspace}
\def\DeltaACP{\ensuremath{(+5.7\pm 2.4)\%}\xspace}
\def\ACPSymb{\ensuremath{{\cal A}_\CP}\xspace}
\def\ArawSymb{\ensuremath{{\cal A}_\text{raw}}\xspace}
\def\Acppi{\ensuremath{{\cal A}_{\CP}(\Lbpi)}\xspace}
\def\AcpK{\ensuremath{{\cal A}_{\CP}(\LbK)}\xspace}
\def\Acph{\ensuremath{{\cal A}_{\CP}(\Lbh)}\xspace}
\def\Arawpi{\ensuremath{\ArawSymb(\Lbpi)}\xspace}
\def\ArawK{\ensuremath{\ArawSymb(\LbK)}\xspace}
\def\Arawh{\ensuremath{\ArawSymb(\Lbh)}\xspace}
\def\myKstar{{\ensuremath{\Kbar^{*}(892)^{0}}}\xspace}%
\def\BdjKs{\mbox{\ensuremath{\Bdb\to\jpsi\myKstar}}\xspace}%
\def\AcpB{\ensuremath{{\cal A}_{\CP}(\BdjKs)}\xspace}
\def\ArawB{\ensuremath{\ArawSymb(\BdjKs)}\xspace}
\def\DeltaACPSymb{\ensuremath{\Delta \ACPSymb}\xspace}
\def\ACPBdSyst{\ensuremath{-1.1\pm0.3\%}\xspace} 
\def\ACPDetSyst{\ensuremath{0.0\pm0.8\%}\xspace}
\def\ACPFitSyst{\ensuremath{0.0\pm0.7\%}\xspace}
 
\def\ACPCorrTot{\ensuremath{-1.1}\xspace}
\def\ACPSystTot{\ensuremath{\pm1.2}\xspace}
\newcommand{\DeltaACPTot}[1][E]{\ensuremath{(+5.7\pm 2.4\STAT[#1]\ACPSystTot\SYST[#1])\%}\xspace}
\def\DeltaACPSignif{\ensuremath{2.2\sigma}\xspace}
\def\hadron{\ensuremath{h^-}\xspace}
\def\hadronp{\ensuremath{h^+}\xspace}

\def\Psippi{\mbox{\ensuremath{\jpsi\proton\pim}}\xspace}
\def\PsipK{\mbox{\ensuremath{\jpsi\proton\Km}}\xspace}

\def\Psippim{\mbox{\ensuremath{\jpsi\proton\pim}}\xspace}
\def\PsipKm{\mbox{\ensuremath{\jpsi\proton\Km}}\xspace}

\def\Psippip{\mbox{\ensuremath{\jpsi\antiproton\pip}}\xspace}
\def\PsipKp{\mbox{\ensuremath{\jpsi\antiproton\Kp}}\xspace}

\def\Lbpi{\decay{\Lb}{\Psippi}}
\def\LbK{\decay{\Lb}{\PsipK}}

\def\Lbh{\decay{\Lb}{\jpsi\proton\hadron}}
\def\Lbpim{\decay{\Lb}{\Psippim}}
\def\LbKm{\decay{\Lb}{\PsipKm}}

\def\Lbbpip{\decay{\Lbbar}{\Psippip}}
\def\LbbKp{\decay{\Lbbar}{\PsipKp}}

\def\LbL{\decay{\Lb}{\jpsi\Lz}}

\newcommand{\skipit}[1]{}%
\def\nw{0.23\textwidth}%

\begin{document}

\renewcommand{\thefootnote}{\fnsymbol{footnote}}
\setcounter{footnote}{1}


\begin{titlepage}
\pagenumbering{roman}

\vspace*{-1.5cm}
\centerline{\large EUROPEAN ORGANIZATION FOR NUCLEAR RESEARCH (CERN)}
\vspace*{1.5cm}
\hspace*{-0.5cm}
\begin{tabular*}{\linewidth}{lc@{\extracolsep{\fill}}r}
\ifthenelse{\boolean{pdflatex}}
{\vspace*{-2.7cm}\mbox{\!\!\!\includegraphics[width=.14\textwidth]{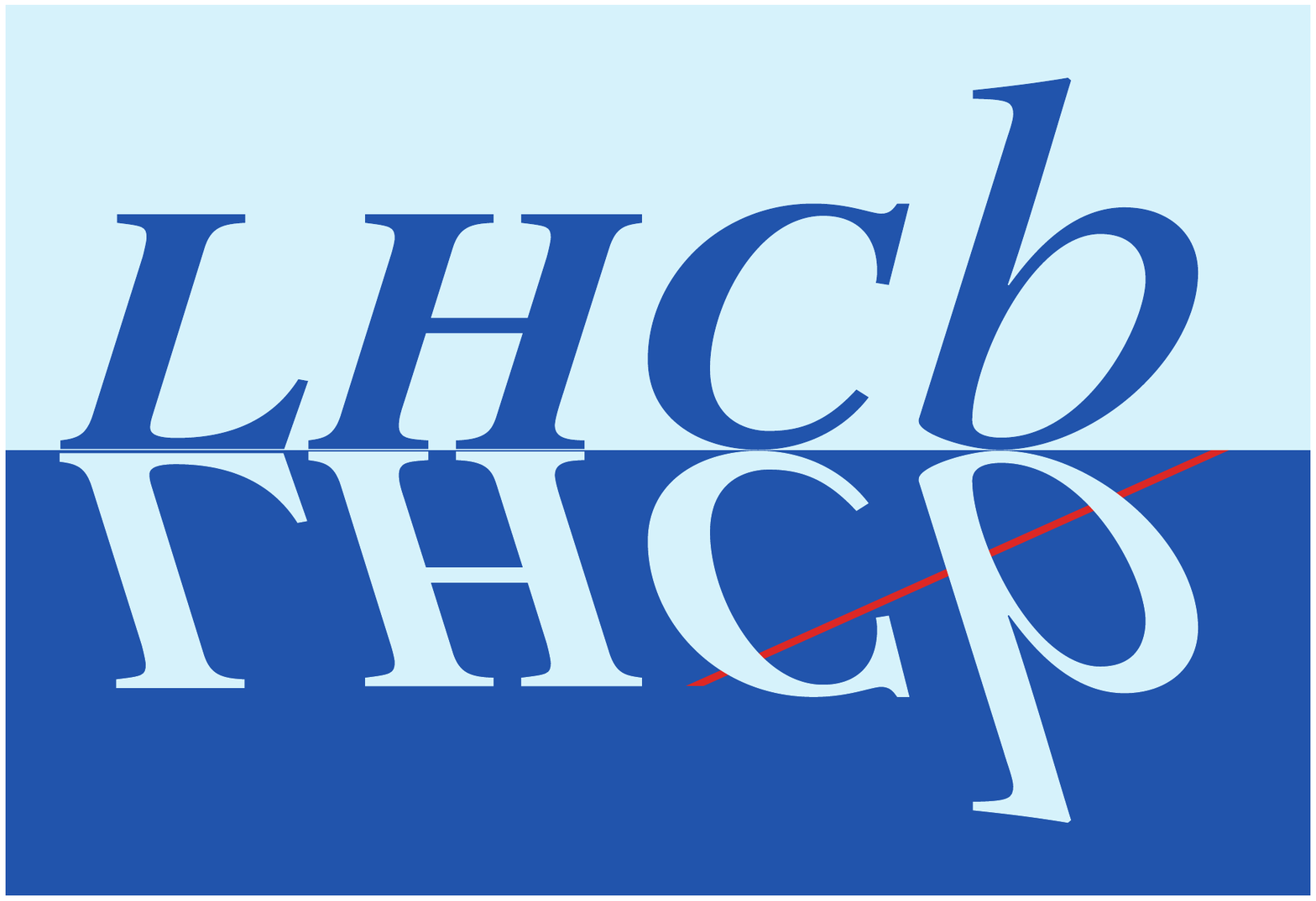}} & &}%
{\vspace*{-1.2cm}\mbox{\!\!\!\includegraphics[width=.12\textwidth]{lhcb-logo.eps}} & &}%
\\
 & & CERN-PH-EP-2014-115 \\  
 & & LHCb-PAPER-2014-020 \\  
 & & 3 June 2014 \\ 
 & & \\
\end{tabular*}

\vspace*{2.0cm}

{\bf\boldmath\huge
\begin{center}
  Observation of the \Lbpi decay
\end{center}
}

\vspace*{1.0cm}

\begin{center}
The LHCb collaboration\footnote{Authors are listed on the following pages.}
\end{center}

\vspace{\fill}

\begin{abstract}
  \noindent
The first observation of the Cabibbo-suppressed decay $\Lambda_b^0\rightarrow J/\psi p \pi^-$ is reported using a data sample of proton-proton collisions at 7 and 8 TeV, corresponding to an integrated luminosity of 3 $\rm fb^{-1}$. A prominent signal is observed and the branching fraction relative to the decay mode $\Lambda_b^0\rightarrow J/\psi p K^-$ is determined to be $$ \frac{{\cal B}(\Lambda_b^0\rightarrow J/\psi p \pi^-)}{{\cal B}(\Lambda_b^0\rightarrow J/\psi p K^-)}=0.0824\pm0.0025\:(\text{stat})\pm0.0042\:(\text{syst}). $$ A search for direct CP violation is performed. The difference in the CP asymmetries between these two decays is found to be $$ {\cal A}_{CP}(\Lambda_b^0\rightarrow J/\psi p \pi^-)-{\cal A}_{CP}(\Lambda_b^0\rightarrow J/\psi p K^-)=(+5.7\pm 2.4\:(\text{stat})\pm1.2\:(\text{syst}))\%, $$ which is compatible with CP symmetry at the $2.2\sigma$ level. 
\end{abstract}
\vspace*{2.0cm}
\begin{center}
  Published in JHEP 07 (2014) 103
\end{center}

\vspace{\fill}

{\footnotesize 
\centerline{\copyright~CERN on behalf of the \lhcb collaboration, 
license \href{http://creativecommons.org/licenses/by/3.0/}{CC-BY-3.0}.}}
\vspace*{2mm}

\end{titlepage}


\newpage
\setcounter{page}{2}
\mbox{~}
\newpage

\centerline{\large\bf LHCb collaboration}
\begin{flushleft}
\small
R.~Aaij$^{41}$, 
B.~Adeva$^{37}$, 
M.~Adinolfi$^{46}$, 
A.~Affolder$^{52}$, 
Z.~Ajaltouni$^{5}$, 
S.~Akar$^{6}$, 
J.~Albrecht$^{9}$, 
F.~Alessio$^{38}$, 
M.~Alexander$^{51}$, 
S.~Ali$^{41}$
% ...

{\footnotesize \it
% ...
$ ^{5}$Clermont Universit\'{e}, Universit\'{e} Blaise Pascal, CNRS/IN2P3, LPC, Clermont-Ferrand, France\\
$ ^{6}$CPPM, Aix-Marseille Universit\'{e}, CNRS/IN2P3, Marseille, France\\
% ...
$ ^{9}$Fakult\"{a}t Physik, Technische Universit\"{a}t Dortmund, Dortmund, Germany\\
% ...
$ ^{37}$Universidad de Santiago de Compostela, Santiago de Compostela, Spain\\
$ ^{38}$European Organization for Nuclear Research (CERN), Geneva, Switzerland\\
% ...
$ ^{41}$Nikhef National Institute for Subatomic Physics, Amsterdam, The Netherlands\\
% ...
$ ^{46}$H.H. Wills Physics Laboratory, University of Bristol, Bristol, United Kingdom\\
% ...
$ ^{51}$School of Physics and Astronomy, University of Glasgow, Glasgow, United Kingdom\\
$ ^{52}$Oliver Lodge Laboratory, University of Liverpool, Liverpool, United Kingdom\\
% ...
}
\end{flushleft}

\cleardoublepage

\
\renewcommand{\thefootnote}{\arabic{footnote}}
\setcounter{footnote}{0}



\pagestyle{plain} 
\setcounter{page}{1}
\pagenumbering{arabic}


%


\section{Introduction}\label{sec:Introduction}
The study of \bquark-baryon decays is of considerable interest both to probe the dynamics of heavy 
flavour decay processes and to search for the effects of physics beyond the Standard Model. 
Owing to their non-zero spin, \bquark baryons provide the potential to improve the limited 
understanding of the helicity structure of the underlying Hamiltonian~\cite{Mannel:1997pc,*Hiller:2007ur}.

Beauty baryons are copiously produced at the LHC, where the \Lb baryon cross-section is
about half of the size of the \Bz meson production in the forward 
region~\cite{LHCb-PAPER-2014-004,*LHCb-PAPER-2011-018}.
The ATLAS, CMS and LHCb collaborations
measured the \Lb lifetime~\cite{LHCb-PAPER-2014-003,LHCb-PAPER-2013-065,*Chatrchyan:2013sxa,Aad:2012bpa},
and the masses of the ground~\cite{LHCb-PAPER-2012-048,Aad:2012bpa} and first excited 
states~\cite{LHCb-PAPER-2012-012}. 
The \Lb polarisation has been measured and found to be 
compatible with zero~\cite{LHCb-PAPER-2012-057}.
The LHCb collaboration has studied 
\Lb decays to charmonium \cite{LHCb-PAPER-2012-057,LHCb-PAPER-2013-032},
open charm~\cite{LHCb-PAPER-2013-056,LHCb-PAPER-2014-002},
charmless states~\cite{LHCb-PAPER-2013-061} and final states induced by
electroweak penguins~\cite{LHCb-PAPER-2013-025}. 
No evidence for \CP violation has been reported in decays of baryons. 
Searches with \bquark-baryon decays have been performed with the decay channels 
\decay{\Lb}{\proton{}\pim}, 
\proton{}\Km~\cite{Aaltonen:2014vra} and \KS{}\proton{}\pim~\cite{LHCb-PAPER-2013-061}.
The corresponding theoretical literature is still 
limited compared to that on \B meson decays.

The study of \decay{\bquark}{\cquark{}\cquarkbar{}\quark} decays can be used to constrain 
penguin pollution in the determination of the \CP-violating phases in \Bd
and \Bs mixing~\cite{DeBruyn:2010hh,*Faller:2008gt}.
While decays originating from the 
\decay{\bquark}{\cquark{}\cquarkbar{}\squark} transitions, such as \LbL or \LbK,
are largely dominated by the tree amplitudes, penguins amplitudes are
enhanced in Cabibbo-suppressed \decay{\bquark}{\cquark{}\cquarkbar{}\dquark} transitions, 
such as the \Lbpi decay.

This article reports the first observation of the \Lbpi decay and the determination of 
its branching fraction relative to the Cabibbo-favoured mode \LbK.
The latter, which was recently observed, has been used to obtain a precise measurement 
of the ratio of \Lb to \Bz lifetimes~\cite{LHCb-PAPER-2013-032,LHCb-PAPER-2014-003}. 
Its absolute branching ratio is yet to be determined.
A measurement of the \CP asymmetry difference between the \Lbpi and \LbK decays
is also reported. The analysis is based on a data sample
of proton-proton collisions, corresponding to an integrated luminosity of 
1\:\invfb at a centre-of-mass energy of 7\:\tev and 2\:\invfb at 8\:\tev, 
collected with the LHCb detector.

\section{Detector and software}
\label{sec:Detector}
The \lhcb detector~\cite{Alves:2008zz} is a single-arm forward
spectrometer covering the \mbox{pseudorapidity} range $2<\eta <5$,
designed for the study of particles containing \bquark or \cquark
quarks. The detector includes a high-precision tracking system
consisting of a silicon-strip vertex detector surrounding the proton-proton
interaction region, a large-area silicon-strip detector located
upstream of a dipole magnet with a bending power of about
$4{\rm\,Tm}$, and three stations of silicon-strip detectors and straw
drift tubes~\cite{LHCb-DP-2013-003} placed downstream of the magnet.
The combined tracking system provides a momentum measurement with
a relative uncertainty that varies from 0.4\% at low momentum to 0.6\% at 100\gevc,
and an impact parameter measurement with a resolution of 20\mum for
charged particles with large transverse momentum, \pt. 
Different types of charged hadrons are distinguished using information
from two ring-imaging Cherenkov (RICH) detectors~\cite{LHCb-DP-2012-003}. 
Photon, electron and hadron candidates are identified by a calorimeter 
system consisting of
scintillating-pad and preshower detectors, an electromagnetic
calorimeter and a hadronic calorimeter. Muons are identified by a
system composed of alternating layers of iron and multiwire
proportional chambers~\cite{LHCb-DP-2012-002}.
The trigger~\cite{LHCb-DP-2012-004} consists of a
hardware stage, based on information from the calorimeter and muon
systems, followed by a software stage, which applies a full event
reconstruction.

Candidate events are first required to pass the hardware trigger,
which selects muons with $\pt>1.48\gevc$. 
In the subsequent software trigger, at least
one of the candidate muons is required to be inconsistent  
with originating from any primary interaction.
Finally, the muon pair is required to
form a vertex that is significantly
displaced from all primary vertices (PV) and to 
have a mass within 120\:\mevcc of the known 
\jpsi mass.

In the simulation, proton-proton collisions are generated using
\pythia~\cite{Sjostrand:2006za,*Sjostrand:2007gs} with a specific \lhcb
configuration~\cite{LHCb-PROC-2010-056}.  Decays of hadronic particles
are described by \evtgen~\cite{Lange:2001uf}, in which final state
radiation is generated using \photos~\cite{Golonka:2005pn}. The
interaction of the generated particles with the detector and its
response are implemented using the \geant
toolkit~\cite{Allison:2006ve, *Agostinelli:2002hh} as described in
Ref.~\cite{LHCb-PROC-2011-006}.

\section{Event Selection}\label{Sec:Method}\label{Sec:Selection}
The \Lbpi and \LbK decays are reconstructed with the \jpsi decaying to two muons. 
Charge conjugation is implied throughout except in the definition of the \CP asymmetry.

Candidate \decay{\jpsi}{\mumu} decays are reconstructed from oppositely charged particles passing loose
muon-identification requirements and with $\pt>500\:\mevc$. They
are required to form a good quality vertex and have a mass in the range
$[3030,3150]\:\mevcc$.
This interval corresponds to about eight times the $\mumu$ mass resolution at the \jpsi mass
and covers part of the \jpsi meson radiative tail. 

Candidate \Lb baryons are selected from combinations of \jpsi{} candidates and two
oppositely charged particles, one of which must be compatible with the proton hypothesis.
The proton candidate is required to have a momentum, $p$, larger than 5\:\gevc,
while the second charged particle must have $p>3\:\gevc$.
Both particles must have $\pt>500\:\mevc$ and be inconsistent with coming from any PV.
All four charged particles are required to be consistent with coming from 
a common vertex.

The reconstructed mass and decay time of the \Lb candidates
are obtained from a kinematic fit~\cite{Hulsbergen:2005pu} that
constrains the mass of the \mumu pairs to the known 
\jpsi mass and the \Lb candidate to originate from the PV.
If the event has multiple PVs, all combinations are considered.
Candidates are required to have a reconstructed decay time larger than $0.2\:\ps$.

To remove backgrounds from \LbL decays, candidates that have a \proton{}\pim mass
within 5\:\mevcc of the \Lz baryon mass are vetoed. To remove reflections from
\decay{\Bs}{\jpsi{}\Pphi} decays, candidates are also vetoed if the hadron-pair 
mass is less than 1035\:\mevcc 
when applying a \Kp mass hypothesis to both particles.

The remaining candidates are split into samples of \Lbpi and \LbK 
according to the estimated probabilities that the charged meson
candidate is a kaon or a pion. 
These probabilities are determined
using a neural network (NN) exploiting information from the RICH detectors,
calorimeter and muon systems, as well as track quality.
Particles with a larger pion probability
are treated as \Lbpi candidates, otherwise they are treated as \LbK candidates.
In addition, the larger of these two probabilities is required to be in excess of 5\%.
The \Lb candidates are required to be in the mass range $[4900,6100]\,\mevcc$.
After this selection $4.3\times10^5$ \Lbpi and $1.9\times10^5$ \LbK candidates remain.

The selection described above is not sufficient to isolate the small \Lbpi signal from 
the combinatorial background.
The initial selection is therefore followed by a multivariate analysis,
based on another NN \cite{0402093v1}.
The NN classifier's output is used as the final selection variable.

The NN is trained entirely on data, using the \LbK signal as a proxy for the 
\Lbpi decay.
The training is performed using half of the \LbK candidates chosen at random.
The other half is used to define the normalisation sample, allowing
an unbiased measurement of the \LbK yield.
The training uses signal and background weights determined using the \sPlot~technique
\cite{Pivk:2004ty} and obtained by performing a maximum likelihood fit
to the unbinned mass distribution of the candidates meeting the loose selection criteria.

The fit probability density function (PDF) is defined as the sum of the \Lb signal component
and the combinatorial background components. 
The parameterisation of the individual components is described in the next section.

Several reflections from \Bd, \Bs and \Lb decays, reconstructed using misidentified particles,
must be accounted for in the mass spectrum.
In order to avoid the training being biased by these reflections, 
all candidates that have a mass compatible with the 
\Bd, \Bs or \Lb mass after swapping the \proton and \kaon assignments
with any of \pion, \kaon, or \proton are removed from the training.

The NN classifier uses information about the candidate kinematic distributions, 
vertex and track quality, 
impact parameter and particle identification information on the proton. 
The most discriminating quantities are the proton particle identification probability, the kinematic fit 
quality and the kaon separation of the PV
in this order.
The variables that are used in the NN are chosen to avoid correlations 
with the reconstructed \Lb mass
and to have identical distributions in \Lbpi and \LbK simulated data.

Final selection requirements of the NN classifier output are chosen
to optimise the expected statistical precision on the \Lbpi signal yield.
The expected signal and background yields entering the sensitivity estimation
are obtained from the training sample
by scaling the number of surviving \LbK candidates by the expected yield based on 
an assumed branching fraction ratio of $0.1$.
The expected background is extrapolated from the 
number of \Lbpi candidates in the mass range $[5770, 6100]\:\mevcc$.
After applying the final requirement on the NN classifier output,
the multivariate selection rejects 99\%  of the background
while keeping 75\% of the \LbK signal, relative to the initial selection.

After applying the full selection, about $0.1\%$ of the 
selected events have more than one candidate sharing at least one track,
or more than one PV that can be used to determine the kinematic properties 
of the candidate. 
In these cases one of the candidates or PVs is used at random.

\section{Signal and background description}\label{Sec:Fit}

For the candidates passing the NN requirements, the yields of \Lbpi and \LbK decays are determined
from unbinned maximum likelihood fits to the mass distributions of 
reconstructed \Lb candidates.
The PDF is defined as the sum of a \Lb signal component, a combinatorial background
and the sum of several reflections. 

The signal shape is parametrised by a Gaussian distribution with
power-law tails on both sides, as indicated by simulation.
The parameters describing the tails are taken from simulation,
while the mean and width of the Gaussian are allowed to vary in the fit.
The combinatorial background contribution is described by an exponential function,
with yield and slope parameter allowed to vary freely.

\begin{figure}[t]
\def\nw{0.5\textwidth}
\includegraphics[width=\nw]{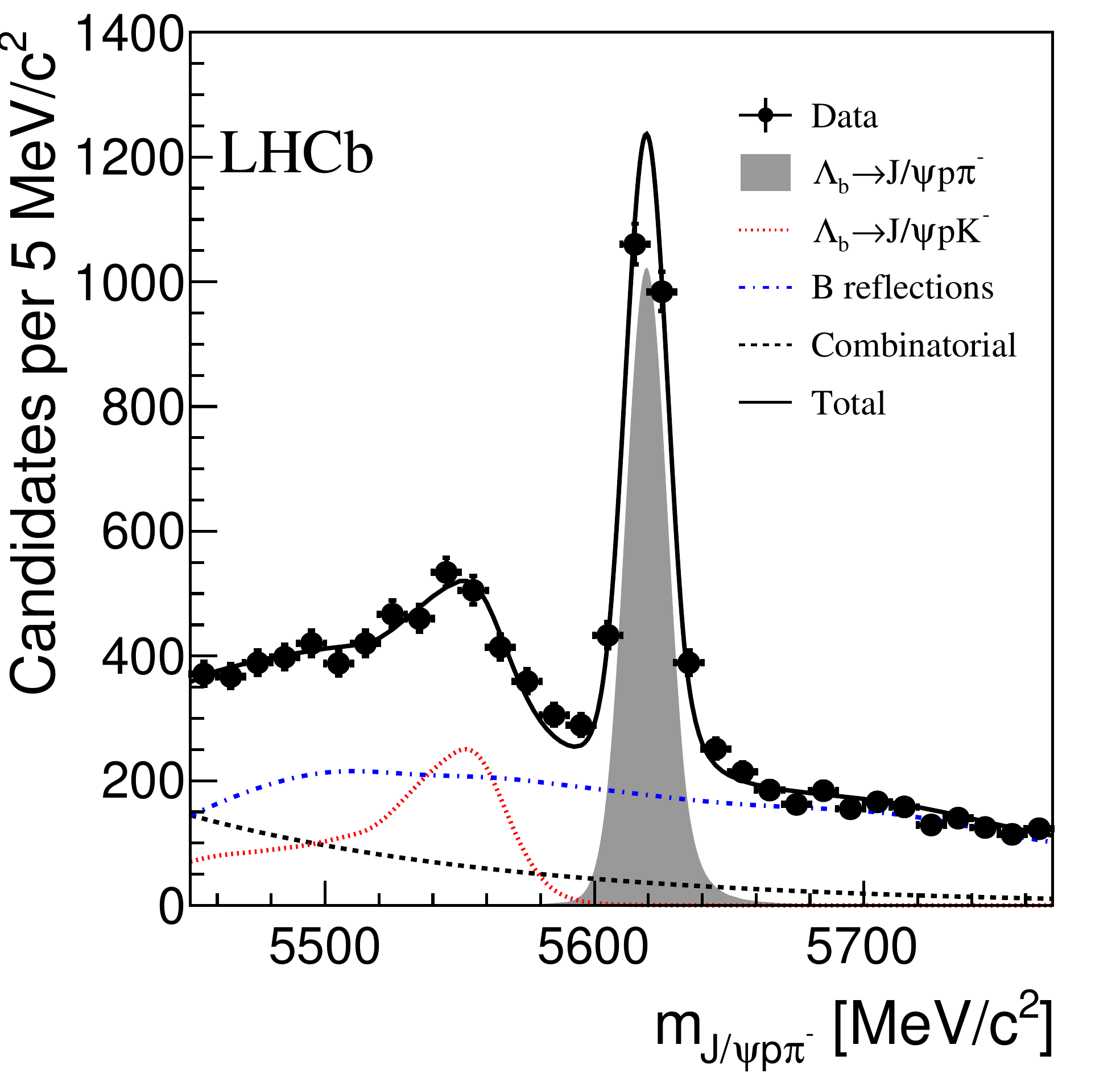}
\includegraphics[width=\nw]{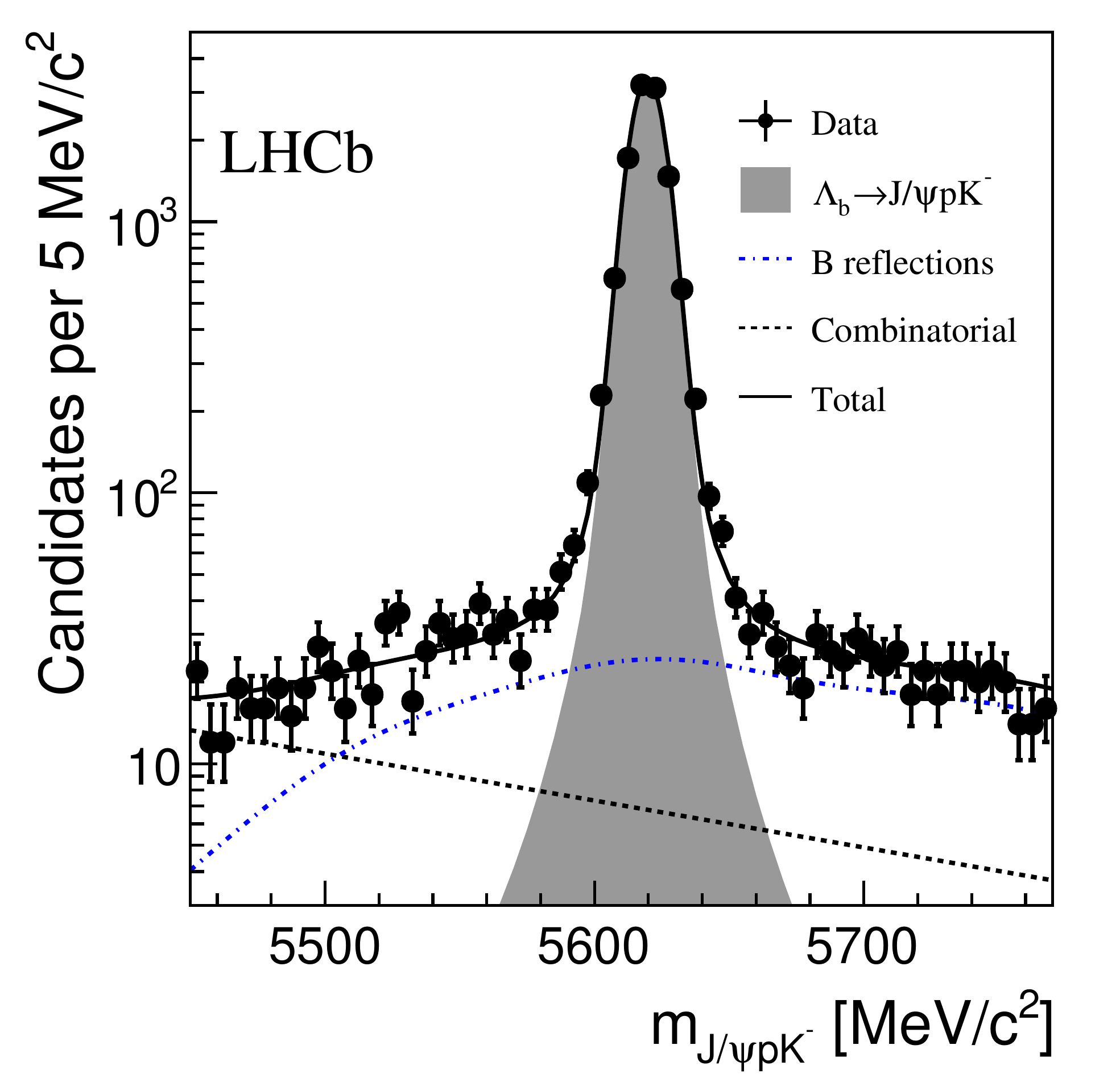}
\caption{Distribution of (left) \Psippi and (right) \PsipK masses
  with fit projections overlaid.
  For \LbK candidates, only the normalisation sample is shown.}
\label{Fig:Fit}
\end{figure}

Several peaking backgrounds due to decays of \bquark hadrons to \jpsi mesons 
and two charged hadrons,
where one or both hadrons are misidentified, survive the selection. 
In this fit they are not vetoed, unlike in the training of the NN,
except for candidates consistent with the \decay{\Bs}{\jpsi\Pphi} hypothesis.
Instead, their contribution is modelled by smoothed non-parametric functions
determined from simulated data. The respective yields are determined by swapping the mass 
assignments of the \proton, \pion and \kaon in turn and searching for 
peaks at the \Bd, \Bs or \Lb masses.
In the \Lbpi fit, significant backgrounds are found from the decays
\LbK (with \Km identified as \pim),
\decay{\Bd}{\jpsi{}\Kp{}\pim} (with \Kp identified as \proton),
and \decay{\Bs}{\jpsi{}\Kp{}\Km} (with one \kaon identified as \proton and the other as \pion).
In the \LbK fit, the main contributions are from the decays
\decay{\Bdb}{\jpsi{}\pip{}\Km} (with \pip identified as \proton),
\decay{\Bs}{\jpsi{}\Kp{}\Km} (with \Kp identified as \proton), and
\LbbKp (with \Km and \proton swapped).
These yields are then used as Gaussian constraints on the normalisation of the 
reflection background shapes.
The results of the fits are shown in Fig.~\ref{Fig:Fit}. The contributions
of the reflections are summed, except for the large \LbK reflection in the \Lbpi
fit, which is shown separately.

Low-mass contributions of partially reconstructed 
\Lbpi{}\piz and \LbK{}\piz decays, where the \piz is not 
considered in the combination, are investigated.
Adding such a component to the fit, with a mass shape taken from simulation, 
results in yields compatible with zero and does not change the signal yields.

In total $11\,179\pm109$ \LbK and $2102\pm61$ \Lbpi decays are obtained.
The \LbK yield, having been determined on the half of the data not used in the training, is
multiplied by two, resulting in a ratio of \Lbpi to \LbK yields of \RawRatio.

The shapes used in the mass fit are varied to determine 
a systematic uncertainty related to the mass 
model. No significant differences are found when trying alternate signal parameterisation that still
result in a good fit. Changing the peaking backgrounds PDF, or letting
their yield free in the fit, change their relative fit fractions, as well
as that of the combinatorial background, but does not affect the signal yields.
The combinatorial background model is changed from an exponential to a second-order
polynomial, which results in a \Lbpi (\LbK) yield reduced by 2.1\% (0.3\%). These 
variations are added in quadrature and used to estimate a systematic uncertainty 
on the ratio of branching fractions.

\section[\CP asymmetry]{\bf\boldmath\CP asymmetry}\label{Sec:CP}
The same fit procedure is repeated separately for baryon (tagged by a positively charged proton)
and antibaryon candidates. All parameters of the fits are determined again separately,
except for the signal shape and the combinatorial background slope, which 
are taken from the fit to all candidates. 
A total of 
$1131\pm40$ \Lbpim, $964\pm38$ \Lbbpip, $5655\pm77$ \LbKm and $5529\pm76$ \LbbKp decays 
are found,
corresponding to the raw asymmetries 
\begin{align}
  \Arawpi & \equiv \frac{N(\Lbpim)-N(\Lbbpip)}{N(\Lbpim)+N(\Lbbpip)} \\
          &=  \RawLbpiCP, \nonumber\\
  \ArawK & =  \RawLbKCP.\nonumber
\end{align}

The procedure to assess the systematic uncertainties related to the shape of the mass distribution,
described in Sec.~\ref{Sec:Fit},
is repeated to determine the sensitivity of the raw asymmetries. A total variation of 0.7\% 
is obtained, which is dominated by a change in \Arawpi when using a second-order 
polynomial background model.

The raw decay-rate asymmetry can be decomposed as
\begin{align}\label{eq:CP:pi}
\Arawh & =  \Acph + {\cal A}_\text{prod}(\Lb) - 
                                {\cal A}_\text{reco}(\hadronp) + {\cal A}_\text{reco}(\proton),
\end{align}
where the terms on the right hand side of the equation are the \CP-violating, 
\Lb production and reconstruction asymmetries of the hadron $h^{\pm}=\pipm,\Kpm$ and the proton, 
respectively. Reconstruction asymmetries are defined following the convention
${\cal A}_\text{reco}(\hadronp)\equiv\frac{\epsilon(\hadronp)-\epsilon(\hadron)}{\epsilon(\hadronp)+\epsilon(\hadron)}$
throughout, where $\epsilon$ is the reconstruction efficiency.
The production asymmetry ${\cal A}_\text{prod}$ and the proton reconstruction
asymmetry cancel in the difference of the
two asymmetries
\begin{align}\label{eq:CP:Delta2}
\Delta {\cal A}_{\CP} & \equiv   \Acppi  - \AcpK \nonumber \\
        & =  \Arawpi - \ArawK + {\cal A}_\text{reco}(\pip) - {\cal A}_\text{reco}(\Kp).
\end{align}

The kaon and pion asymmetries can be determined from the 
raw asymmetry of the \BdjKs decay with \decay{\myKstar}{\Km{}\pip}.
It has been measured~\cite{LHCb-PAPER-2012-021} as 
\begin{equation}\label{eq:CP:Kstarzb} 
  \ArawB \equiv \frac{N(\Bdb)-N(\Bd)}{N(\Bdb)+N(\Bd)} = (-1.10\pm0.32\pm0.06)\%,
\end{equation}
where the first uncertainty is statistical and the second systematic.
It can be decomposed as 
\begin{align}\label{eq:CP:K2}
\ArawB  & =  \AcpB - \kappa{\cal A}_\text{prod}(\Bd)\\&  + {\cal A}_\text{reco}(\pip) - {\cal A}_\text{reco}(\Kp)\nonumber \\
        & \approx  {\cal A}_\text{reco}(\pip) - {\cal A}_\text{reco}(\Kp),
\end{align}
where $\kappa$ is a dilution factor due to \Bd mixing and ${\cal A}_\text{prod}(\Bd)$ is the \Bd
production asymmetry, which is compatible with zero~\cite{LHCb-PAPER-2013-018}. 
Under the assumption of no \CP asymmetry in the \BdjKs decay and negligible production
asymmetry, this value can thus be taken as the combined kaon and pion reconstruction asymmetry,
and is consistent with measurements in other decay 
modes to kaon and pions~\cite{LHCb-PAPER-2013-018,LHCb-PAPER-2014-013}. 

The difference of \CP asymmetries in the \Lbpi and \LbK decays can then be rewritten as
\begin{align}\label{eq:CP:Delta}
\DeltaACPSymb  & =  \Arawpi - \ArawK + \ArawB \\&= \DeltaACP\nonumber,
\end{align}
where the uncertainty is statistical only.

The kaon and pion momenta in \Lb decays are not identical to those in \BdjKs decays,
which could induce different detector asymmetries in the \Lb and \Bd modes. 
This is investigated by weighting the 
\Lbpi and \LbK data to match the pion and kaon momentum
distributions observed in \BdjKs decays.
The value of \DeltaACPSymb changes by 0.8\%, which is assigned as the systematic uncertainty
related to reconstruction asymmetries.

Local \CP asymmetries in the Dalitz plane are also searched
for using the technique outlined in Ref.~\cite{Bediaga:2012tm,*Bediaga:2009tr}.
No significant local asymmetries are found.

\section{Efficiency corrections and systematic uncertainties}\label{Sec:Dalitz}
The raw quantities need to be corrected to determine the physics quantities.
The efficiency of the selection requirements is studied with simulation. 
Some quantities are known not to be well reproduced in simulation, namely
the \Lb transverse momentum and lifetime, the particle multiplicity, 
and the \Lbpi and \LbK decay kinematic properties. For all these quantities the simulated data are
weighted to match the observed distributions in data. They are obtained with the 
\sPlot\ technique using 
the \Lb candidate mass as the control variable. 

For three-body \bquark-hadron decays, both the signal decays and the dominant combinatorial 
backgrounds populate regions close to the kinematic boundaries of the  
\Psippi and \PsipK Dalitz plot~\cite{Dalitz:1953cp}. 
For more accurate modelling of these regions, it is convenient to transform the conventional 
Dalitz space to a rectangular 
space (hereafter referred to as the square Dalitz plot~\cite{Aubert:2005sk}). We follow 
the procedure described in Ref.~\cite{LHCb-PAPER-2013-061}.

The \Lbpi and \LbK decays have different detector acceptance, reconstruction
and selection efficiencies. They are determined from simulated data, which are
weighted to match the experimental data.
The main differences are induced by 
\begin{enumerate}[i.]
\item the detector acceptance, as the efficiency of
  \LbK is 6\% larger than that for the \Lbpi decays due to the lower kinetic energy release in the former,
  which causes smaller opening angles;
\item the reconstruction and preselection efficiency, which is 4\% larger in 
  \Lbpi decays due to the 
  average total and transverse momentum of the final state particles being larger than in \LbK decays;
\item the particle identification requirements on the \pim or \Km, 
  which are more efficient on \Lbpi decays by 7\%;
\item the \Pphi veto, which removes 7\% (3\%) of the \LbK (\Psippi) signal.
\end{enumerate}
Overall, these effects result in a further correction on the ratio of branching fractions of 
\Lbpi to \LbK decays of \BFSystMC. 

The efficiency of particle identification is not perfectly modelled in simulation.
The kaon and pion identification efficiencies are further weighted after the kinematic weighting 
described above, using a large sample of \Dstar-tagged \Dz\to\Km{}\pip decays. 
The uncertainties are determined by varying the weights of the simulated data within their uncertainties,
yielding a correction of \BFSystPID. 
The kinematic properties of the proton
in the \Lbpi and \LbK decays are found to be the same. The same applies to the muons.
The efficiencies of proton and muon identification therefore 
cancel in the ratio of branching fractions, as well as in the \CP asymmetries.

The trigger efficiency is determined using simulation, which is validated using \LbK decays from data. 
Differences
between the \Lbpi and \LbK decays efficiencies are at the percent level and are assigned as 
systematic uncertainties.

The value of the \Lb lifetime used in simulation is taken from Ref.~\cite{PDG2012}, 
and is 3\% lower than the current most precise measurement~\cite{LHCb-PAPER-2014-003}.
The simulated data is weighted to account for this and the difference 
is assigned as a systematic uncertainty.

\label{Sec:Syst}

\begin{table}[tb]\centering
\caption{Corrections and related systematic uncertainties on 
  the ratio of \Lbpi to \LbK branching fractions (BF) and on the 
  difference between the \CP asymmetries \DeltaACPSymb.
  The corrections are multiplicative on the branching fraction and additive 
  for the asymmetry.}\label{Tab:Syst}
\begin{tabular}{lCR}
Source       & \text{BF} &  \multicolumn{1}{C}{\DeltaACPSymb}  \\
\hline
Simulation-based corrections & \BFSystMC      &  \multicolumn{1}{c}{-}  \\
PID          & \BFSystPID       &  \multicolumn{1}{c}{-}  \\
Trigger      & \BFSystTrigger   &  \multicolumn{1}{c}{-}  \\
\Lb lifetime & \BFSystLifetime & \multicolumn{1}{c}{-} \\
Mass distribution model   & \BFSystFit         & \ACPFitSyst \\
\BdjKs           &   - & \ACPBdSyst     \\
Detection asymmetries &   - & \ACPDetSyst \\
\hline
Total        & \BFSystTot       & \ACPCorrTot\ACPSystTot\%  \\
\end{tabular}
\end{table}

The estimates of the systematic uncertainties described above
are summarised in Table~\ref{Tab:Syst}, along with the 
total obtained by summing them in quadrature.
The uncertainty on the 
ratio of branching fractions is dominated by the uncertainty on
corrections obtained from simulation,
mostly due to the unknown decay kinematic properties of the \Lbpi decay.
For \DeltaACPSymb, the mass model distribution and the detection asymmetries contribute about equally.


\section{Results and Conclusions}\label{Sec:Results}
A signal of the Cabibbo-suppressed \Lbpi decay is observed for the first time
using a data sample of proton-proton collisions at 7 and 8\:\tev, corresponding to an integrated 
luminosity of 3\:\invfb.
Applying the appropriate corrections detailed in Table~\ref{Tab:Syst}, 
the ratio of branching fractions is measured to be 
\begin{align*}
\frac{{\cal B}(\Lbpi)}{{\cal B}(\LbK)} 
                                       & =  \BFCorrected.
\end{align*}

Assuming these decays are dominated by tree \decay{\bquark}{\cquark{}\cquarkbar{}\dquark}
and \decay{\bquark}{\cquark{}\cquarkbar{}\squark} transitions, respectively, the 
ratio of Cabibbo-Kobayashi-Maskawa (CKM) 
matrix elements $|\Vcd|^2/|\Vcs|^2$ times ratio of phase space factors
is approximately $0.08$,
compatible with the measured value. This ratio can also be compared to that
of the decays \decay{\Lb}{\Lc\Dm} and \decay{\Lb}{\Lc\Ds}, which involve the same 
quark lines as \Lbpi and \LbK, respectively. The ratio of \decay{\Lb}{\Lc\Dm} and \decay{\Lb}{\Lc\Ds}
has been measured as
$0.042\pm0.003\stat\pm0.003\syst$~\cite{LHCb-PAPER-2014-002}, which is 
consistent with this measurement, when taking into account the 
\Dm and \Ds meson decay constants~\cite{PDG2012} and the different ratio of
phase space factors.

\begin{figure}[tb]
\def\nw{0.5\textwidth}
\includegraphics[width=\nw]{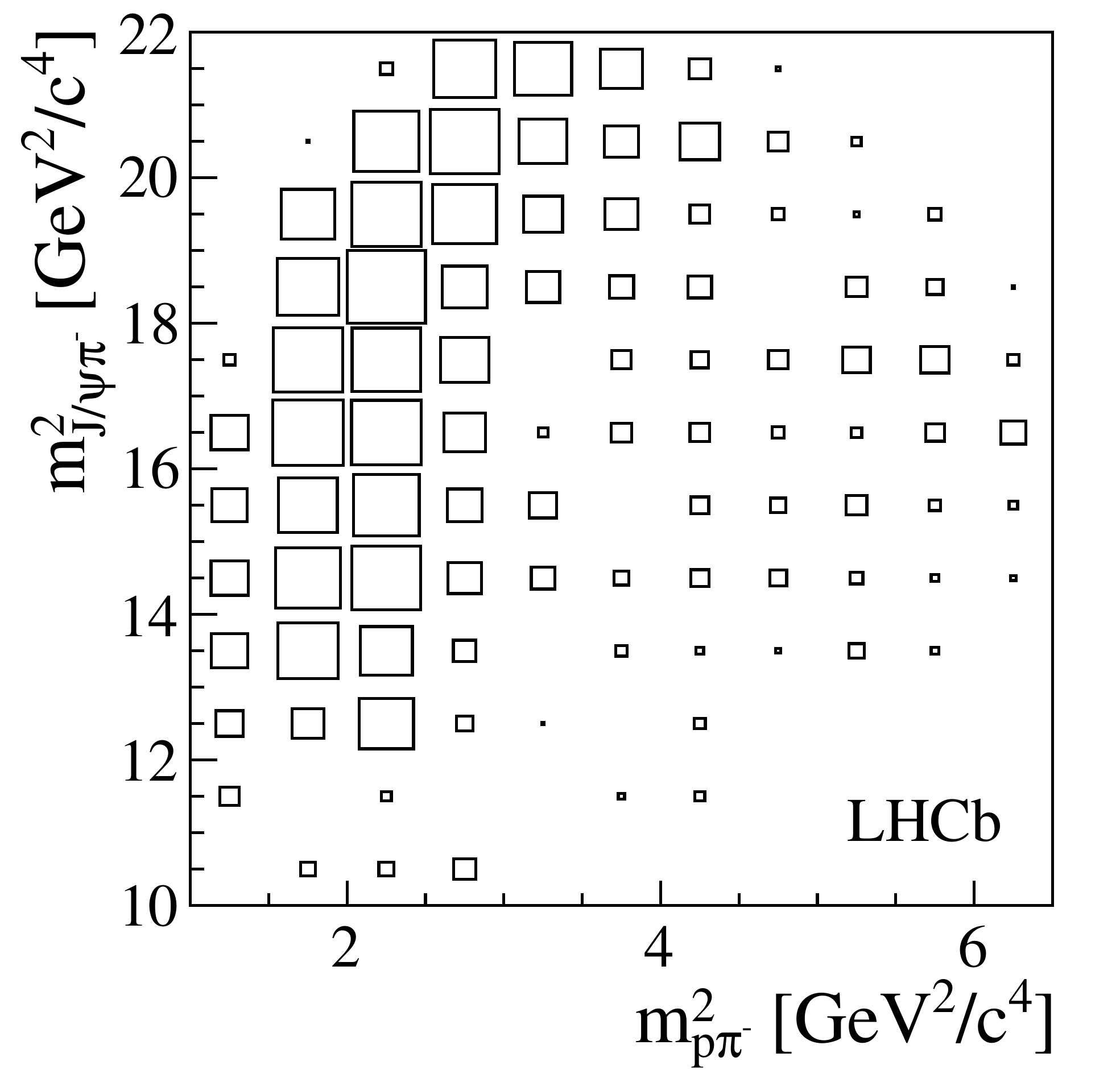}
\includegraphics[width=\nw]{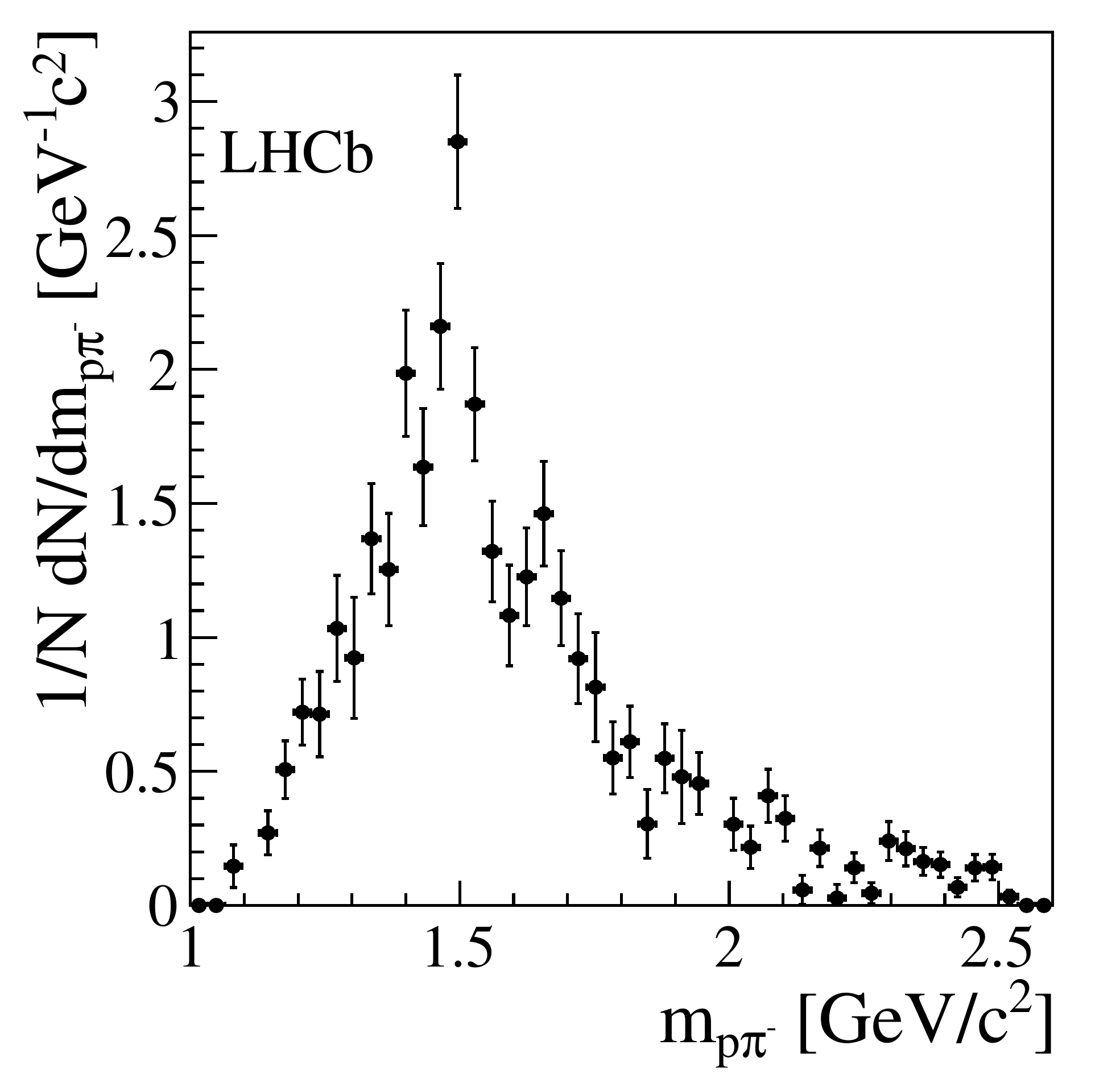}\\
\includegraphics[width=\nw]{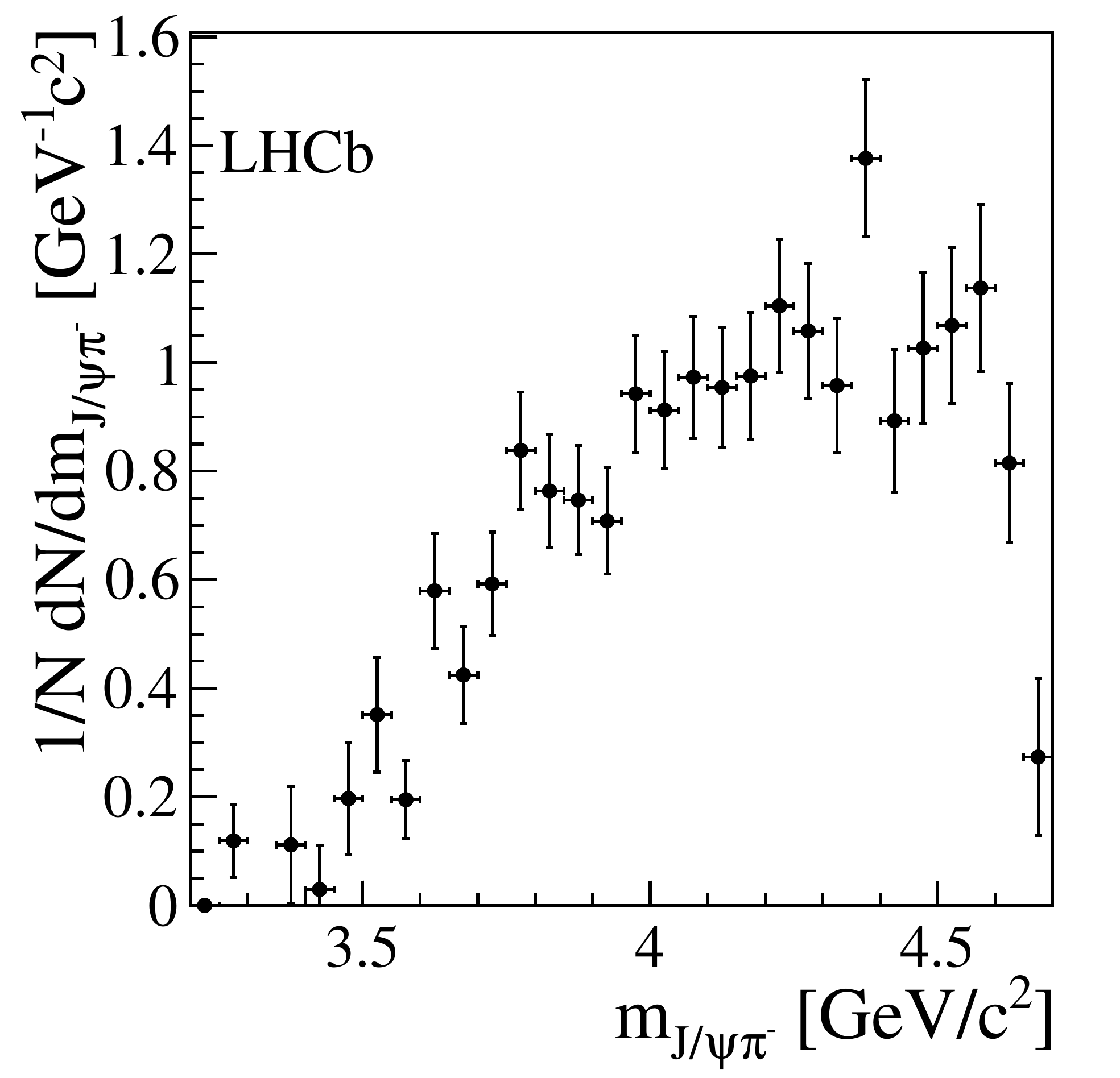}
\includegraphics[width=\nw]{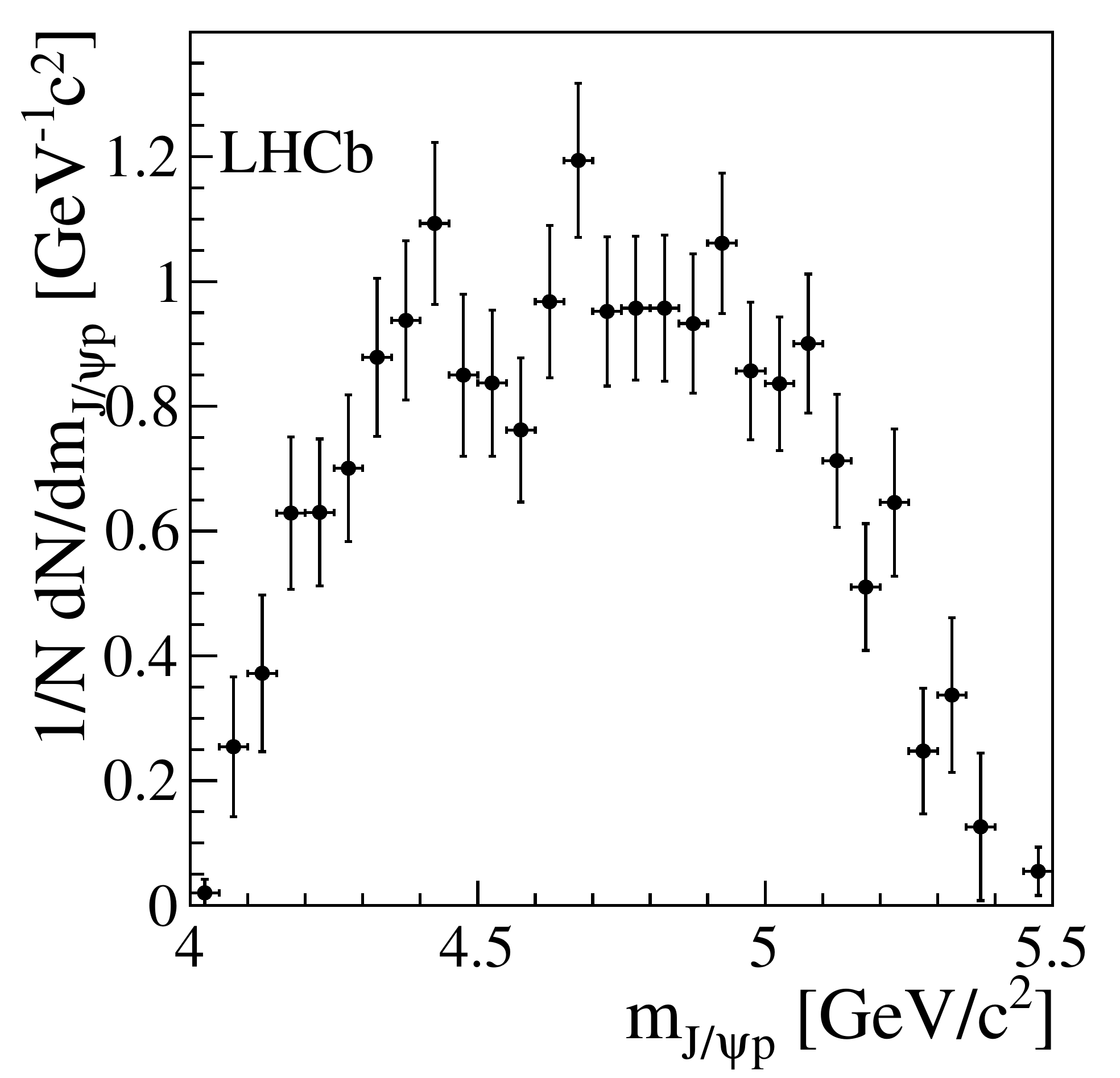}\\
\caption{Efficiency-corrected and background-subtracted 
  \Lbpi Dalitz plane and projections normalised to unit area 
  on the \proton{}\pim,
  \jpsi{}\pim and \jpsi{}\proton axes. }
\label{Fig:Dalitz}%
\end{figure}

Background-subtracted and 
efficiency-corrected distributions of kinematic
distributions determined in the \Lbpi
decay are shown in Fig.~\ref{Fig:Dalitz}. In this case the 
\Lb mass is fixed to its known value 
and the kinematic properties
recomputed~\cite{Hulsbergen:2005pu}.
No attempt is made to fit the decay rate on 
the Dalitz plane. The \proton{}\pim mass distribution shows a rich resonant structure, as
expected from fits to fixed-target experiment data~\cite{Anisovich:2011fc,Arndt:2006bf},
and suggests the presence of the narrow $N(1520)$ or $N(1535)$, the $N(1650)$, 
as well as the broad $N(1440)$ resonances. No signs of exotic structures in the \jpsi{}\pim
or \jpsi{}\proton mass distributions are seen.
More data and further studies will be needed to inverstigate the 
underlying dynamics of this decay.

The \CP asymmetry difference between the \Lbpi and \LbK decays is measured to be  
\begin{equation*}
  \Delta A_{\CP} = A_{\CP}(\Lbpi)-A_{\CP}(\LbK) = \DeltaACPTot,
\end{equation*}
corresponding to a \DeltaACPSignif deviation from zero. No indications
of large local \CP asymmetries in the Dalitz plane are observed.
The precision of these measurements illustrate the potential 
of Cabibbo-suppressed \Lb decays in studies of direct \CP violation.


\section*{Acknowledgements}
\noindent We express our gratitude to our colleagues in the CERN
accelerator departments for the excellent performance of the LHC. We
thank the technical and administrative staff at the LHCb
institutes. We acknowledge support from CERN and from the national
agencies: CAPES, CNPq, FAPERJ and FINEP (Brazil); NSFC (China);
CNRS/IN2P3 (France); BMBF, DFG, HGF and MPG (Germany); SFI (Ireland); INFN (Italy); 
FOM and NWO (The Netherlands); MNiSW and NCN (Poland); MEN/IFA (Romania); 
MinES and FANO (Russia); MinECo (Spain); SNSF and SER (Switzerland); 
NASU (Ukraine); STFC (United Kingdom); NSF (USA).
The Tier1 computing centres are supported by IN2P3 (France), KIT and BMBF 
(Germany), INFN (Italy), NWO and SURF (The Netherlands), PIC (Spain), GridPP 
(United Kingdom).
We are indebted to the communities behind the multiple open 
source software packages on which we depend. We are also thankful for the 
computing resources and the access to software R\&D tools provided by Yandex LLC (Russia).
Individual groups or members have received support from 
EPLANET, Marie Sk\l{}odowska-Curie Actions and ERC (European Union), 
Conseil g\'{e}n\'{e}ral de Haute-Savoie, Labex ENIGMASS and OCEVU, 
R\'{e}gion Auvergne (France), RFBR (Russia), XuntaGal and GENCAT (Spain), Royal Society and Royal
Commission for the Exhibition of 1851 (United Kingdom).



\addcontentsline{toc}{section}{References}
\setboolean{inbibliography}{true}
\bibliographystyle{LHCb}
\bibliography{main,LHCb-PAPER,LHCb-CONF,LHCb-DP,MyBib}

\end{document}